\documentclass[twocolumn,prl,amsmath,amssymb,aps,superscriptaddress]{revtex4-1}
\usepackage{graphicx}
\usepackage{dcolumn}
\usepackage{bm}

\usepackage{xcolor}

\newcommand{\ve}[1]{\ensuremath{\mbox{\boldmath$#1$}}}

\begin{document}
\title{Flow Navigation by Smart Microswimmers via Reinforcement Learning}
\author{Simona Colabrese}
\affiliation{Department of Physics and INFN, University of Rome Tor Vergata, Via della Ricerca Scientifica 1, 00133 Rome, Italy}
\email{simona.colabrese@roma2.infn.it}

\author{Kristian Gustavsson}
\affiliation{Department of Physics and INFN, University of Rome Tor Vergata, Via della Ricerca Scientifica 1, 00133 Rome, Italy}
\affiliation{Department of Physics, University of Gothenburg,  Origov\"agen 6 B, 41296 G\"oteborg, Sweden}
\author{Antonio Celani}
\affiliation{Quantitative Life Sciences, The Abdus Salam International Centre for Theoretical Physics, Strada Costiera 11, 34151 Trieste, Italy}

\author{Luca Biferale}
\affiliation{Department of Physics and INFN, University of Rome Tor Vergata, Via della Ricerca Scientifica 1, 00133 Rome, Italy}
\date{\today}

\begin{abstract}
Smart active particles can acquire some limited knowledge of the fluid environment from simple mechanical cues and exert a control on their preferred steering direction. Their goal is to learn the best way to navigate by exploiting the underlying flow whenever possible. As an example, we focus our attention on smart gravitactic swimmers. These are active particles whose task is to reach the highest altitude within some time horizon, given the constraints enforced by fluid mechanics. By means of numerical experiments, we show that swimmers indeed learn nearly optimal strategies just by experience. A reinforcement learning algorithm allows particles to learn effective strategies even in difficult situations when, in the absence of control, they would end up being trapped by flow structures. These strategies are highly nontrivial and cannot be easily guessed in advance. This Letter illustrates the potential of reinforcement learning algorithms to model adaptive behavior in complex flows and paves the way towards the engineering of smart microswimmers that solve difficult navigation problems.\footnote{\textbf{Postprint version of the article published on Phys. Rev. Lett. 118, 158004 (2017) DOI: 10.1103/PhysRevLett.118.158004}}
\end{abstract}
\maketitle
Swimming microorganisms can take advantage of environmental stimuli to bias their motility patterns in order to achieve some biologically relevant goal, some examples being chemotaxis, phototaxis, and gravitaxis~\cite{pedley1992hydrodynamic,fenchel2002microbial,kiorboe2001marine}. Taking inspiration from nature, artificial micro- and nanoswimmers with active internal or external controls could be engineered to execute specialized tasks in complex environments~\cite{lauga2009hydrodynamics,ebbens2010pursuit,ghosh2009controlled,mair2011highly,fischer2011magnetically,wang2012nano,gazzola2014reinforcement,gazzola2016learning}. These tasks could be, for example, exploiting advection by the flow to reach specific regions, enhancing transport and mixing, or escaping from potentially dangerous hydrodynamical fluctuations \cite{michalec2015turbulence,tanyeri2010hydrodynamic,genin2005swimming,zirbel2000reversible,sengupta2016phytoplankton}. Here, the general questions that we want to address are:
can these smart particles learn how to escape their hydrodynamical fate just by sensing simple environmental cues and by reacting to these with the modification of a few control parameters of their dynamics? Is learning also feasible in complex flows, which unavoidably lead to poor performances in the absence of control? What do good strategies look like? Could they be easily intuited {\it a priori}?
To what extent can strategies that perform well in a given environment provide an advantage also under other conditions?

In this Letter, we advocate for the use of reinforcement learning as a general framework to construct efficient strategies for microscopic motility and to train smart particles to accomplish long-term tasks.
Reinforcement learning is based on the prolonged and continued interaction between agent and environment, during which the agent---here, a particle or microorganism---learns how to behave optimally by trial and error \cite{book:sutton}. The great potential of this approach has been recently demonstrated in the very complex navigation task of thermal soaring in large-scale turbulent environments \cite{reddy2016learning}. Here, we show by means of numerical experiments that it can be successfully applied to the microscopic problem of gravitaxis in a flowing fluid.
\noindent
We consider active particles that swim with constant speed while being carried away by the underlying flow. The direction of the swimming velocity is  determined by the competition between a stabilizing torque that tries to align the particle with a preferred swimming direction---one might think about it as the orientation of a rudder---and the rotation induced by the flow vorticity which could favor or oppose this alignment.
If the particle has some control on the preferred direction, how should it operate to achieve its goal, that is, to obtain, in the long run, the largest possible progression in the upward direction? In a quiescent fluid, the optimal choice for the preferred direction is to steadily point upwards. This is realized in {\it gyrotactic particles} by means of an uneven distribution of mass, see for example \cite{kessler1985hydrodynamic}. In the presence of an underlying flow, this strategy may reveal to be highly ineffective \cite{santamaria2014gyrotactic,durham2009disruption,durham2013turbulence}. Indeed, naive gyrotactic particles in a steady flow with horizontal vortex rolls can aggregate in tight clusters and remain trapped at a given height (see Ref. \cite{durham2011gyrotaxis} and Fig. \ref{fig:1} below). \\
{\it Smart gravitactic particles}, on the contrary, are endowed with the
ability of obtaining some partial information about the regions of the flow that they are visiting. They can use this knowledge to choose directions that maximize the total ascent in the long run, which allows them to  escape trapping regions and seek ``elevator'' regions of the flow. This might be seen as a basic implementation of more realistic behaviour as the one given by some species of phytoplankton that are able to actively reorganize its internal organelles in response to fluid mechanical cues \cite{sengupta2016phytoplankton}. Reinforcement learning provides a way to construct these efficient strategies just by accumulating experience.\\
%%%%%%%%%%%%%%%%%%%%%%%%%%%%%%%%%%%%%%%%%%%%%%%%
\noindent {\it Gyrotactic swimmers.}---We consider pointlike, neutrally buoyant particles that are small enough for inertial effects to be ignored. The flow is not affected by the particles.
%fig 1 %%%%%%%%%%%%%%%%%%%%%%%%%%%%%%
\begin{figure}[tbp]
	\centering
	\includegraphics*[scale=0.3,trim={8mm 6mm 18mm 2mm}]{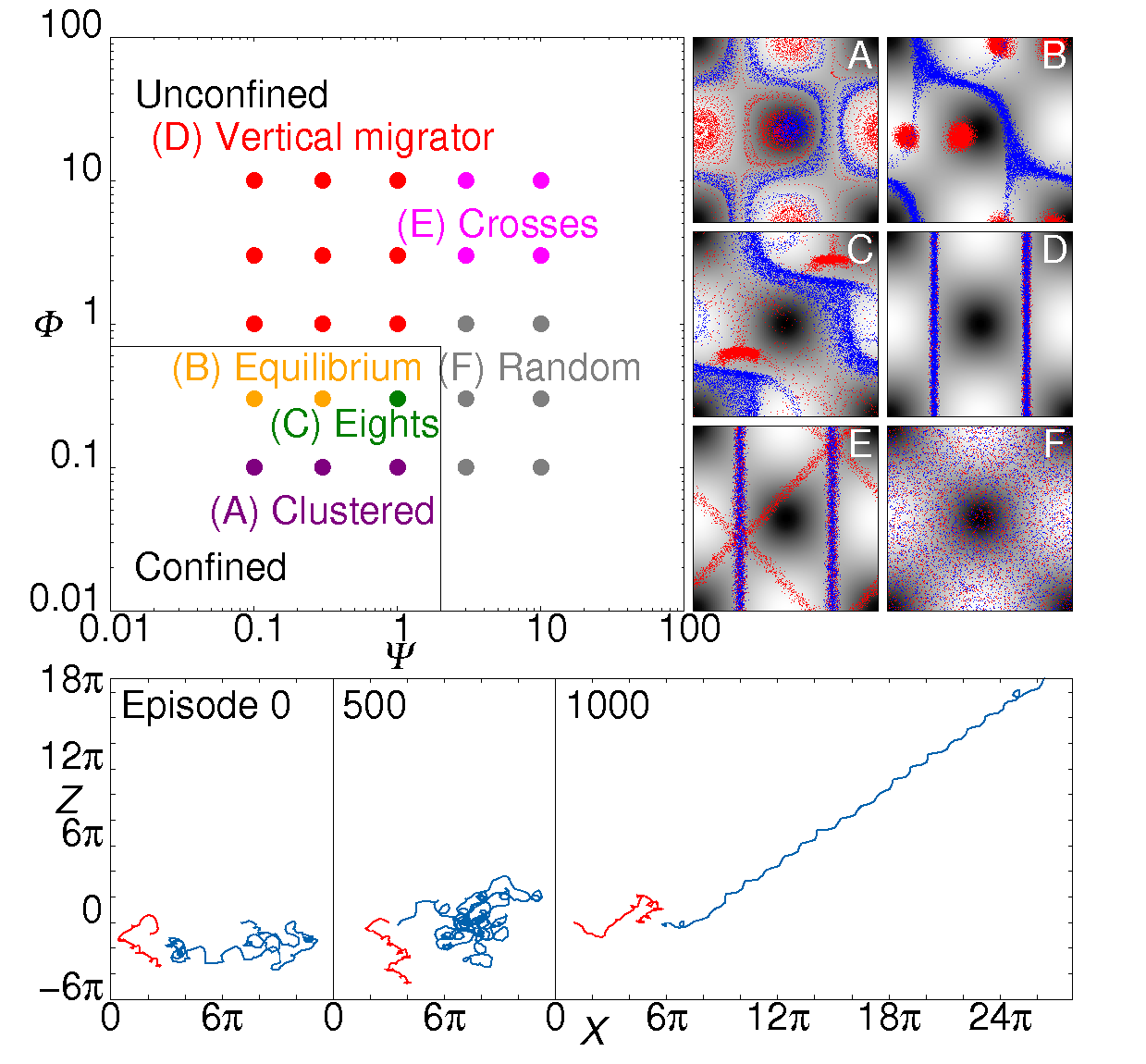}
	\caption{%
		Top left: phase diagram of gyrotactic particles in a Taylor-Green vortex flow.  Each circle represents one of the $25$ investigated parameter pairs. Six distinct patterns emerge: confined  (a)--(c) and unconfined (d)--(f).
		Top right: gyrotactic trajectories (red) for each of the 6 patterns plotted on a periodic domain. For the sake of representation, unconfined trajectories are reinjected in the basic $2\pi \times 2\pi$ domain.  Trajectories for smart gravitactic particles, after learning the optimal policy to choose the preferred direction ${\bf k}_a$,	are shown in blue.
		The vorticity field is shown in grey scale.
		Bottom: a set of representative trajectories at different learning episodes for smart gravitactic particles (blue) compared with typical  trajectories for naive gyrotactic particles (red) confined in a
		trapping dynamics (case $C$ above).
	}\label{fig:1}
\end{figure}
The trajectories $\ve x(t)$ are determined by a superposition of the fluid velocity $\ve u$ and the swimming velocity $v_s\, \ve{p}$,
\begin{equation}\label{eq:Pos}
\dot{\ve{x}} = \ve{u}  + v_s\, \ve{p} + \sqrt{2D_0}\ve\eta\,.
\end{equation}
Here, $v_s$ is a constant speed, $\ve\eta(t)$ is Gaussian white noise,
$\langle \eta_i(t)\eta_j(t') \rangle = \delta_{ij}\delta(t-t')$, and $D_0$ is the translational diffusivity.
The direction of the swimming velocity $\ve{p}$ obeys~\cite{pedley1992hydrodynamic}
\begin{equation}\label{eq:Dir}
\dot{\ve{p}} =  \dfrac{1}{2B}[\ve{k}_a- (\ve{k}_a\cdot \ve{p})\ve{p}] + \dfrac{1}{2}\ve\omega \times \ve{p} +\sqrt{2D_R} \ve\xi,
\end{equation}
where $ \ve{k}_a$ is a unit vector that defines the preferred direction, $B$ is the time scale of alignment,
$\ve\xi(t)$ is a white-in-time Gaussian noise, and $D_R$ is the rotational diffusivity.
The index $a$ runs over a discrete set (the actions),
each corresponding to a possible choice for the preferred direction.
For a flow with a characteristic velocity $u_0$ and vorticity $\omega_0$,  the particle motion is characterized by two dimensionless parameters:
the swimming number, $\Phi = v_s / u_0 $, quantifying the swimming speed relative to the ambient flow and the stability number, $ \Psi = B \omega_0 $,
measuring the strength of the viscous torque exerted by vorticity relative to the stabilizing torque.
The dimensionless translational and rotational diffusivities are chosen to be small (see section 3 in Supplemental Material \cite{SM}).

Naive gyrotactic particles have only one single $\ve{k}_a$ that always points upwards.
 A systematic exploration of the
 parameter space for a gyrotactic particle in a Taylor-Green flow (TGF) 
 made of a periodic array of counterrotating vortices has been performed in Ref.~\cite{durham2011gyrotaxis}, and it has revealed the existence of different regimes. The flow configuration can be considered a model for convection in two dimensions \cite{young1989anomalous,sarracino2016nonlinear} and can be realized in a laboratory with rotating cylinders or in ion solutions in an array of magnets \cite{solomon1988chaotic,tabeling2002two}.
In the top panel of Fig.~\ref{fig:1}, we show the phase diagram for naive gyrotactic particles in the TGF, $ \ve{u} = (u_0/2) [-\cos x \sin z,\sin x \cos z]$.
The results coincide qualitatively with those presented in Ref.~\cite{durham2011gyrotaxis} with the only difference that the noise terms in Eqs.~\eqref{eq:Pos} and~\eqref{eq:Dir} remove the occurrence of strictly periodic orbits and other fragile behaviors.
Notably, the bottom part of phase space ($\Phi \ll 1$), where the flow strongly affects the dynamics, is characterized by a strong reduction in vertical motion either because of confinement (for fast realignment $\Psi \ll 1$) or due to random undirected motion (for slow realignment $\Psi \gg 1$).

Smart gravitactic particles should be able to significantly improve the ascent by appropriately choosing their preferred direction $\ve{k}_a$, according to the environmental cues that they are receiving. Anticipating our results---that will be presented below in full detail---we show in the bottom panel of Fig.~\ref{fig:1} the trajectories of smart particles at different stages of the learning process for a given point (labeled $C$ in the left top panel) in the parameter space.
Evidently, as experience is accumulated, the smart particle performs better and better and eventually achieves a large upward drift. For different values of the parameters, different gains in vertical motion are obtained, but in all cases, we observe at least some improvement. In the top right panel of Fig.~\ref{fig:1} we show a comparison between the spatial distribution patterns of naive gyrotactic and smart gravitactic particles in various regimes. For parameters leading to confinement of naive particles (a)--(c), one can appreciate how smart particles have learned how to concentrate preferentially in regions where the underlying flow facilitates ascent.\\
{\it Learning gravitaxis using smart particles.}---Key ingredients in the reinforcement learning framework are to identify what environmental cues the agent can sense (the states $s$),  what it can do (the actions $a$), and what reinforcement signals it receives in response to its behavior (the rewards $r$) \cite{book:sutton}.
In our setup, particles can perceive only a crude representation of their current swimming direction $\ve{p}$ and of the flow. We choose the set of possible states to be the product of two subsets, one indexing the vorticity level of the underlying flow  (the vorticity is $\omega={\bm \nabla} \times {\bm u}$) and the other one labeling
the instantaneous swimming direction. In short, the discrete state space is $\mathcal{S} = \mathcal{S}_{\omega} \times \mathcal{S}_{p} $, where $\mathcal{S}_{\omega} = \{\omega_-,\omega_0, \omega_+ \}$ are three coarse-grained  vorticity states (negative, close to zero, and positive), and  $\mathcal{S}_{p} = \{\uparrow,
\downarrow, \rightarrow, \leftarrow \}$ are four coarse-grained directions of the instantaneous swimming velocity,
 pointing mainly upwards, downwards, rightwards, or leftwards (see section 1 in the Supplemental Material \cite{SM} for a detailed description).
The set of actions comprises four  preferred swimming directions,
${\ve k}_a \in \mathcal{A} = \{\uparrow, \downarrow, \rightarrow, \leftarrow \} $. Particles evolve according to
Eqs.~\eqref{eq:Pos} and \eqref{eq:Dir} with a given action $a$ that is chosen according to some strategy.
When a particle changes  state, $s_n \to s_{n+1}$, because it has moved into a region with a different vorticity level or its swimming direction points to a different angular sector, a reward $r_{n+1}$ is issued. The reward is given by the net increase in altitude experienced by the particle while being in the old state, $$r_{n+1}= z(s_{n+1})-z(s_n)\,, $$ where $ z(s_n) $ is the initial $z$ coordinate of the particle in state $s_n$. In the new state, a new action is chosen and the cycle is repeated. The final goal is to maximize the expected return, in this case, the average global long-term vertical displacement, $$R_{\mbox{tot}} = \langle \sum_{n=1}^{N_s} r_n \rangle,$$ where the sum extends up to a  horizon $N_s \gg 1$, and the average is over the realizations of the noise in Eqs.~\eqref{eq:Pos} and~\eqref{eq:Dir} and over initial conditions.

Among the many reinforcement learning algorithms that can produce approximately optimal actions, we adopted $Q$ learning. In a nutshell, it constructs an approximation $Q(s,a)$ of the quality matrix, which is the maximal return that can be achieved starting in state $s$ and taking action $a$. Given a state $s_n$ and a current approximation $Q_n$, the action $a_n$ is selected with a bias to favor actions with higher values of $Q_n$. When the state changes, the current estimate is updated on the basis of the reward that has just been received: if it is larger than expected on the basis of $Q_n$, the quality function is increased accordingly, otherwise it is decreased. Iterating this procedure, nearly optimal strategies can be obtained (see section 2 in the Supplemental Material \cite{SM}).

Operationally, we broke the training sessions into subsequences, called episodes $E$, with $E=1,\dots,N_E$, where $N_E$ is the total number of episodes
of each session. The first episode is initialized with an optimistic $Q$; i.e., all entries are equal and very large. This has the effect to encourage exploration and avoid local maxima. Each episode ends after a fixed number of total state changes $N_s$ and is followed by a new episode with a  random restart of the initial position and orientation of the particle. The initial $Q$ of each restarted episode is given by that obtained at the end of the previous episode (see  Supplemental Material \cite{SM}).

%%%%%%%%%%%%%% fig 2 %%%%%%%%%%%%%%%%%%
\begin{figure}[htbp]
\centering
\includegraphics[scale=0.7]{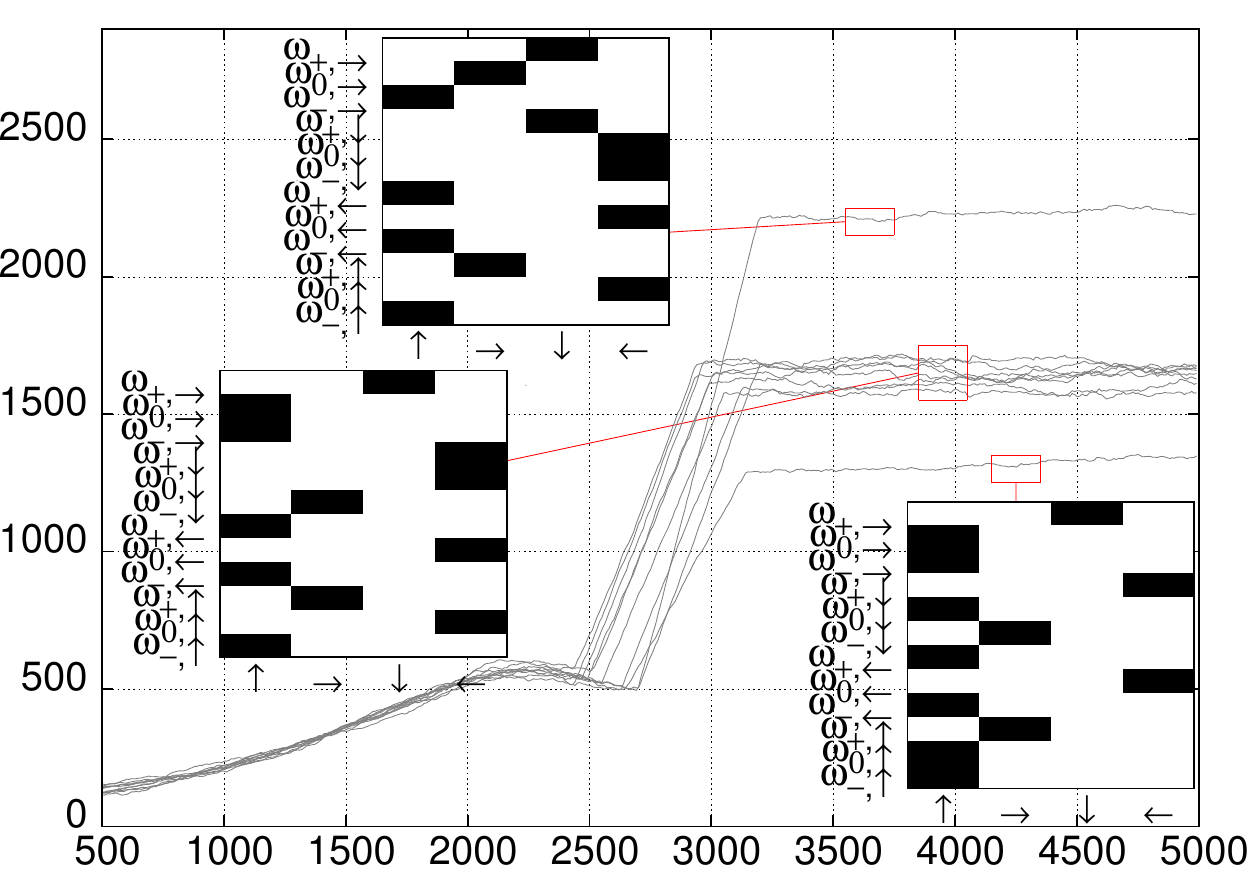}
\caption{%
Dependence of the learning gain $\Sigma(E)$ in percentage vs $E$
for 10 different learning processes (grey curves). Point ($\Psi=1,\Phi=0.3$) region $C$ in Fig.\ref{fig:1}. 
The value of $\Sigma(E)$ that is visualized is averaged locally on a window of 500 episodes.
The insets highlight which preferred directions the smart particle takes for each of the 12 states, according to three final approximately optimal strategies [these are the highest values of the $Q(s,a)$ matrices].
}
\label{fig:3}
\end{figure}

In order to quantify the success of the learning process, we introduce the {\it learning gain} of an episode $\Sigma(E)$.
It measures the relative increase in the return for smart gravitaxis compared to the return for naive gyrotaxis $R_{\mbox{tot},g}$,
\begin{equation}
\Sigma(E) = \frac{R_{\mbox{tot}}}{R_{\mbox{tot},g}}-1\,.
\end{equation}
In Fig.~\ref{fig:3}, we show the evolution of $\Sigma(E)$ for ten different training sessions and for a given choice of parameters $(\Phi,\Psi)$ in regime $C$. We see that the smart particle learns how to improve its performance in a robust way. 
The differences in the asymptotic values across trials are due to the greedy choice of actions based on $Q$ that we have adopted in this particular numerical experiment. Results with better exploratory choices of actions, such as $\epsilon$ greedy, support the same conclusions (see section 3 in the Supplemental Material \cite{SM}).

\begin{figure}[htbp]
\centering
\includegraphics*[scale=0.72,trim={22mm 4mm 19mm 12mm}]{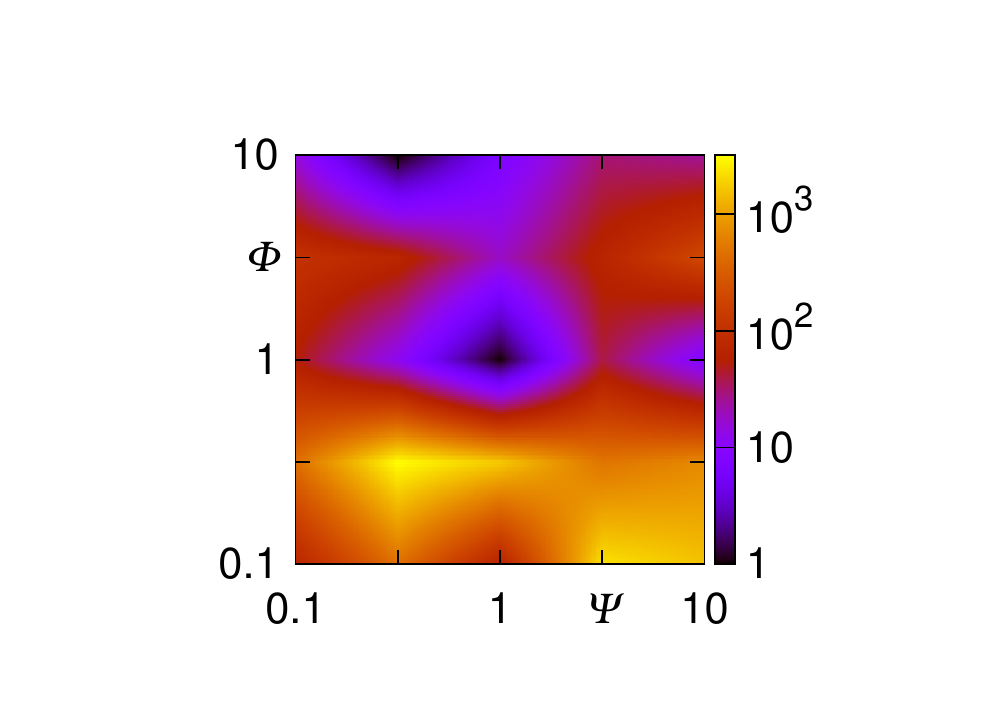}
\includegraphics*[scale=0.43,trim={1mm -9mm 6mm 4mm}]{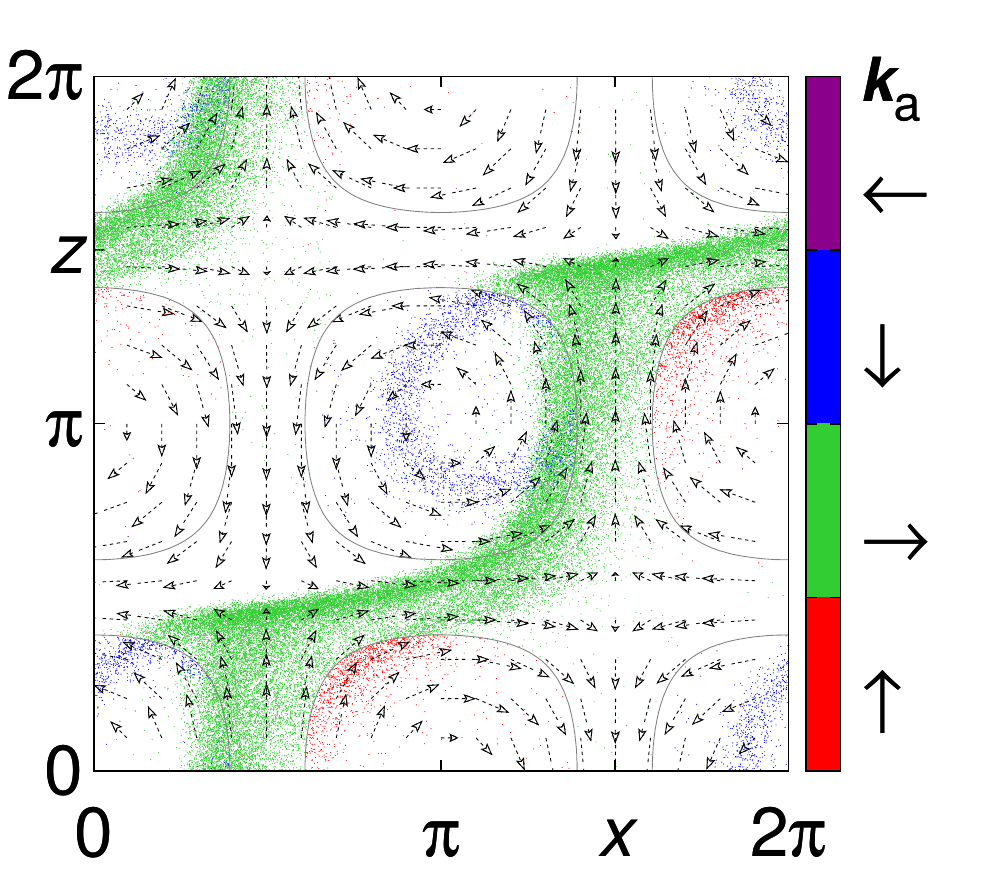}
\caption{Left: final averaged learning gain $\Sigma_{\mbox{avg}}$ (in percentage) for different regions in the parameter space, where the average is made out of $10$ different learning experiments. Right: example of the optimal actions for the smart gravitactic particle that succeeded to escape the confinement, parameters $(\Psi= 1, \Phi=0.3)$ in region $C$ in Fig. \ref{fig:1}. Data obtained from an optimal training using $\epsilon$-greedy exploration. Notice that strategies obtained by permutations that respect the symmetries of the underlying flow would lead to the  same learning gain.}\label{fig:4}
\end{figure}

In the left panel of Fig.~\ref{fig:4}, we present a global overview of the gain for all points in phase space that appear in Fig.~\ref{fig:1}.
When the naive gyrotactic particles are confined or move randomly, the gain is very high, while it is just moderate or low---but always positive---when the naive strategy is already performing well.

It is interesting to notice the nontrivial character of the best strategies, which makes them hard to guess {\it a priori}.
This can be appreciated by visualization of the optimal actions taken in different regions of space (Fig.~\ref{fig:4}, right panel).
We observe that the trajectories of smart gravitactic particles have high density in the up-welling regions, showing that they exploit the ``elevators'' of the flow to reach high altitude.
In particular, when the dynamics bring the particle in a recirculation region, the optimal strategy attempts at steering away in the shortest possible time. Sometimes, this might require a seemingly ineffective choice in the short run, such as pointing downwards; the usefulness of this action can be appreciated only over a long horizon and is therefore difficult to guess in advance.

%%%%%%%%%%%%%%%%%%%%%%%%%

\begin{figure}[htbp]
\centering
\includegraphics*[scale=0.40,trim={24mm 17mm 20mm 8mm}]{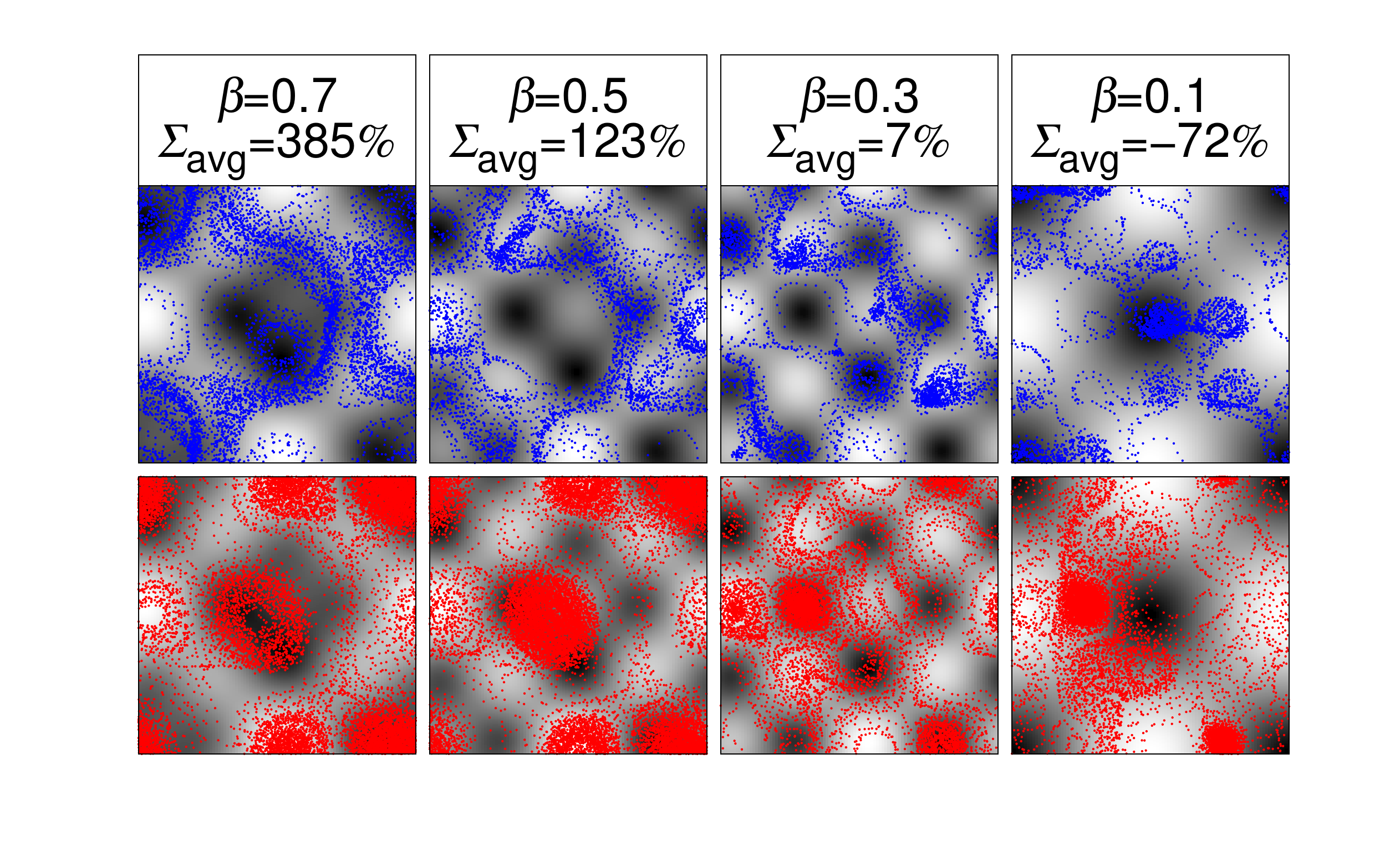}
\caption{Comparison between trajectories of naive gyrotactic (red) and smart gravitactic particles (blue) in a perturbed flow. Smart particles move with a policy obtained for the original flow. Above each column, we put $\beta$ values and the corresponding  $\Sigma_{\mbox{avg}}$. Case $(\Psi = 0.3, \Phi = 0.1) $ region $A$ in Fig. \ref{fig:1}.}\label{fig:1SM}
\end{figure}

{\it Specialized strategies and flow perturbations.}---Given that particles are trained in a specific environment, it is natural to ask how the optimal policy will perform under perturbations of the underlying flow. In general, overspecialized strategies may fail when they have to deal with environments that are wildly different from the ones where the agent has been trained. However, for reasonable classes of environments, they may also display a significant degree of robustness. We addressed this point by evaluating the performance of swimmers trained in the basic Taylor-Green flow
when confronted with a more general flow with vorticity, $\omega(x,z)   = \beta \omega_1 + (1 - \beta)\omega_2$,
where $\omega_1(x,z) =  - u_0 \cos x \cos z$ is the original flow, and
$\omega_2(x,z) =  - u_0 \cos 2(x - \Delta x) \cos 2(y - \Delta y)$ is a rescaled and dephased version of the original flow,  with ($\Delta x, \Delta y) =(3.35,1.83)$.
The parameter $1-\beta$ controls the intensity of the perturbation, with $\beta=1$ corresponding to the original training environment. %In Figure~\ref{fig:1SM} we show the spatial distribution of particles and see that the strategy performs well even down to values $\beta \sim 0.5$ and always outperforms naive gyrotaxis.
In Fig.~\ref{fig:1SM}, we show the spatial distribution of particles and see that the strategy outperforms naive gyrotaxis even down to values $\beta \sim 0.3$, for which the learning gain is of $ 7\%$.
%%%%%%%%%%%%%%%%%%%% fig 1SM %%%%%%%%%%%
%%%%%%%%%%%%%%%%%%%%%%%%%%%%%%%%%%
Moreover, in section 4 of the Supplemental Material \cite{SM}, we also explore the robustness and efficiency of the optimal policy  in presence of time dependent perturbations of the TGF, i.e., even in the case when tracers might have chaotic evolution. We used either the same strategy optimised without the time dependent perturbation, or we allowed the smart particle to learn a new policy. In both cases the smart microswimmers are able to outperform
 the unskilled ones, at least for the explored phase-space points.

{\it Conclusions.}---In this Letter, we have shown how smart particles can learn to accomplish difficult navigation tasks in complex fluid flows. We made no attempt at a fully realistic description of the particle dynamics nor at the actual complexity of real flows, let alone the actual technological implementation of our approach. Rather, our goal was to provide a proof of concept for the possibility to engineer smart microswimmers and to make a case for the use of reinforcement learning algorithms for this purpose. There is enormous room for improvement in many directions: better algorithms, more realistic sensory inputs, and more refined control mechanisms.
For  instance, $Q$ learning algorithms can be implemented to teach particles that can control their relative density with respect to
the underlying fluid to target specific flow configurations. Work in this direction will be reported elsewhere.
We hope that our Letter will spur further research on this field at the interface between fluid mechanics, engineering, and computer science.

%{\it Acknowledgments}---
We acknowledge M. Cencini, G. Reddy and M. Vergassola for useful discussion and for a critical reading of the manuscript.  
S.C. and L.B. acknowledge funding from the European
Research Council under the European Union’s Seventh
Framework Programme, ERC Grant Agreement No.
339032. S.C. acknowledges the hospitality of the Quantitative Life Science research group, The Abdus Salam International Centre for Theoretical Physics, Trieste, Italy. K.G. acknowledges funding from the Knut and Alice Wallenberg Foundation, Dnr.KAW 2014.0048.

%\bibliography{GParticle}
%\bibliographystyle{unsrtplainnat}
%\bibliographystyle{unsrt}

\end{document}